\documentclass[twocolumn,amsmath,amssymb,floatfix,prl]{revtex4-1}

\usepackage{graphicx}

\def\micron{{\ \mu{\rm m}}}					

\def\kHz{{\ {\rm kHz}}}						

\def\us{{\ \mu{\rm s}}}						
\def\ms{{\ {\rm ms}}}						

\def\Rb87{^{87}\rm{Rb}}					

\begin{document}

\title{Phases of a 2D Bose Gas in an Optical Lattice}

\author{K.~Jim\'{e}nez-Garc\'{\i}a$^{1,2}$}
\author{R.~L.~Compton$^1$}
\author{Y.-J.~Lin$^1$}
\author{W.~D.~Phillips$^1$}
\author{J.~V.~Porto$^1$}
\author{I.~B.~Spielman$^1$}
\email{ian.spielman@nist.gov}
\affiliation{$^1$Joint Quantum Institute, National Institute of Standards and Technology, and University of Maryland, Gaithersburg, Maryland, 20899, USA}
\affiliation{$^2$Departamento de F\'{\i}sica, Centro de Investigaci\'{o}n y Estudios Avanzados del Instituto Polit\'{e}cnico Nacional, M\'{e}xico D.F., 07360, M\'{e}xico}

\date{\today}

\begin{abstract}
Ultra-cold atoms in optical lattices realize simple, fundamental models in condensed matter physics. Our $^{87}$Rb Bose-Einstein condensate is confined in a harmonic trapping potential to which we add an optical lattice potential. Here we realize the 2D Bose-Hubbard Hamiltonian and focus on the effects of the harmonic trap, not present in bulk condensed matter systems. By measuring condensate fraction we identify the transition from superfluid to Mott insulator as a function of atom density and lattice depth. Our results are in excellent agreement with the quantum Monte Carlo universal state diagram, suitable for trapped systems, introduced by Rigol {\it et al.} (Phys. Rev. A 79, 053605 (2009)).
\end{abstract}

\maketitle
\begin{figure}[tb]
\begin{center}
\includegraphics[width=3.375in]{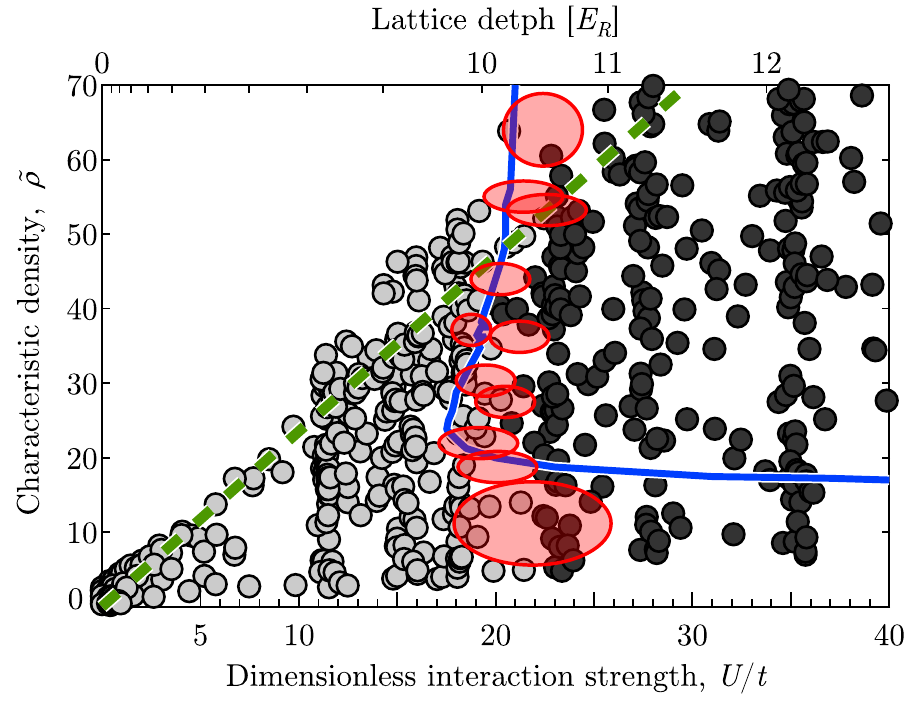}
\end{center}
\caption{Measured state diagram for a 2D Bose gas in the presence of harmonic confinement. The experimental transition boundary (red ovals) was measured from $f(U/t)$ traces like that of Fig.~\ref{CondensateFractionFig} for different $N_{2D}$. Here, the size of the ovals denote the uncertainties in the measurement of $\tilde \rho$ and $U/t$ where we identify the phase transition. The data displayed in Fig.~\ref{CondensateFractionFig} was taken along the green dashed line. The continuous blue line indicates the QMC prediction~\cite{Rigol2009} for the first appearance of a MI in the universal state diagram. The points are colored according to the side of the transition on which they are.}
\label{diagram}
\end{figure}

The Bose-Hubbard (BH) Hamiltonian realized by ultra-cold atoms in optical lattices exemplifies the utility of these systems in studying the idealized lattice models so central to condensed matter physics~\cite{Jaksch1998,Greiner2002,Kohl2005,Spielman2007}. By increasing the depth of the lattice potential, an initially Bose-condensed system can undergo a transition from superfluid (SF) to Mott insulator (MI); careful experiments have pinpointed the critical lattice depth for this transition in 2D~\cite{Spielman2008} and 3D~\cite{Ketterle2007}. The traditional Bose-Hubbard model describes homogeneous systems, but trapped ultra-cold gases are {\it globally} inhomogeneous, potentially containing multiple, spatially separated phases. For sufficiently large systems this inhomogeneity can be understood using the local density approximation (LDA), where each region of the system is treated as being {\it locally} homogeneous. Rigol {\it et al.}~\cite{Rigol2009} introduced a ``universal state diagram'' describing the full configuration of coexisting, spatially separated SF and MI phases in harmonically trapped systems. Here we present measurements of 2D trapped systems and identify the transition from SF to MI as a function of lattice depth and atom number; the resulting experimental state diagram, Fig.~\ref{diagram}, is in good agreement with the quantum Monte Carlo (QMC) predictions of Ref.~\cite{Rigol2009}, which go beyond the LDA.

The BH hamiltonian models the physics of bosons in a lattice potential, here realized with ultra-cold $^{87}$Rb atoms in a 3D optical lattice. The homogeneous BH model includes only pair-wise on-site interactions and nearest-neighbor tunneling, parameterized by an interaction energy $U$ and a tunneling matrix element $t$. At zero temperature this model predicts the existence of a SF phase and MI phases with integer occupation $n=1,2,3...$ per lattice site, determined by the ratio $U/t$ and the chemical potential $\mu$. The importance of interactions is determined by $U/t$ while the density is largely controlled by~$\mu /t$. For weak interactions (small $U/t$) the system is SF, while for $U/t$ larger than a critical value $(U/t)_c$ the system can enter a MI phase. For $U/t\gg(U/t)_c$, the phases alternate between SF and MI, increasing in density as $\mu$ increases~\cite{Fisher1989}.

The homogeneous BH model is not applicable to current trapped, ultra-cold atom experiments, owing to their harmonic trapping potential. The BH Hamiltonian for a lattice with period $d$ superimposed on a symmetric harmonic trap is~\cite{Jaksch1998}
\begin{equation*}\label{BH_hamiltonian}
H=-t\sum_{\langle i,j \rangle}{\hat{b}_{i}^{\dagger}\hat{b}_{j}}+\frac{U}{2} \sum_{i}{\hat{n}_{i}(\hat{n}_{i}-1)}+\sum_{i}{( \epsilon i^{2}-\mu)\hat{n}_{i}}
\end{equation*}
where $\hat{b}_{i}^{\dagger}$ is the creation operator of a boson at site $i$ and $\langle {i},j \rangle$ constrains the sum to nearest neighbor tunneling. The parameter $\epsilon = m \omega ^{2}{d}^2 /{2 } $ describes the harmonic potential where $m$ is the atomic mass and $\nu =2 \pi \omega$ is the trap frequency.

In the LDA, the third term of the Hamiltonian is assumed to be constant over an extended region producing a local chemical potential $\mu_i= \mu - \epsilon i^{2}$; under this approximation the system's properties are computed using the homogeneous BH Hamiltonian. The LDA intuitively explains the evolution of a SF system into a nested collection of alternating SF and MI shells as $U/t$ increases. Such structures were first observed using a magnetic resonance imaging (MRI) approach for a 3D system~\cite{Folling2006a}, by measuring collisional shifts \cite{Campbell2006}, and more recently by direct imaging \cite{Chin2009}.

The inhomogeneity introduced by the trap leads, for sufficiently large $\epsilon$, to a breakdown of the LDA, both quantitatively and qualitatively. For example, the trap potential can increase the critical value $({U/t})_{c}$ at which a Mott insulator first appears~\cite{Rigol2009}. This suggests that a proximity-like effect~\cite{Pannetier2000} stabilizes the SF component where MI was expected.

For a $T=0$ trapped system, the three parameters: $U/t$, $\epsilon$ and the atom number $N$, fully specify the quantum state of the system, even with coexisting regions of SF and MI. Somewhat surprisingly, only two independent variables are sufficient~\cite{Rigol2009}: $U/t$ and a characteristic density $\tilde{\rho}=N \epsilon / t$. By monitoring the dependence of condensate fraction $f$ on~$\tilde{\rho}$ and~$U/t$~\cite{Spielman2008}, we experimentally measure the state diagram for 2D systems in an optical lattice (Fig.~\ref{diagram}).

\begin{figure}[tb]
\begin{center}
\includegraphics[width=3.375in]{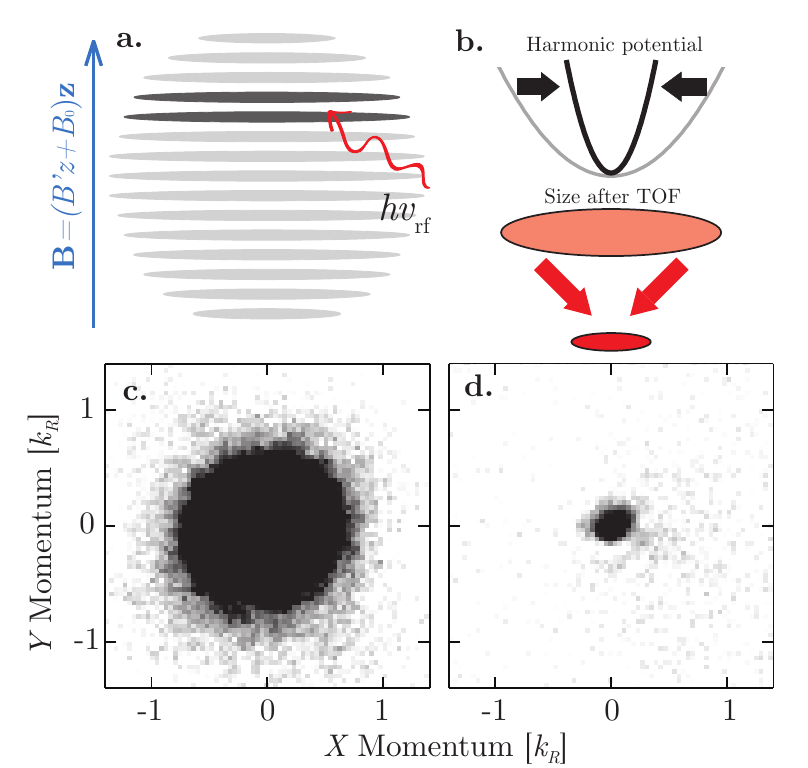}
\end{center}
\caption{a) Our 3D BEC is divided into 2D systems by a deep optical lattice in the presence of a linear magnetic field gradient, both aligned along the $\hat{z}$ direction. We use a MRI technique to selectively address a small subset of 2D systems. b) Matter-wave focusing is used to better resolve the SF phase of the 2D Bose gas; here we show a schematic of the focusing of a single 2D system after free evolution during TOF. c-d) We compare the measured atom distribution (approximating the momentum distribution) of the ensemble of $\approx60$ 2D systems without focusing (c) with that of the addressed 2D systems in the $m_\textrm{F}=0$ sub-level with focusing (d), both with an $\hat{x}$-$\hat{y}$ lattice at 9.5 $E_R$.}
\label{Schem}
\end{figure}

In our experiment we partition a 3D Bose-Einstein condensate (BEC) into an ensemble of nearly independent 2D systems using a 1D optical lattice along $\hat{z}$ (see Fig.~\ref{Schem}). Additional optical lattices along $\hat{x}$ and $\hat{y}$ provide the periodic landscape for the BH Hamiltonian in each 2D system. In this configuration a critical value $(U/t)_c=15.8(20)$~\footnote{Uncertainties reflect the uncorrelated combination of one-standard deviation statistical and systematic uncertainties.}, for which the $n=1$ MI first appears in the ensemble system, was measured in Ref.~\cite{Spielman2008}, at higher temperatures than discussed here. That ensemble measurement, however, could not distinguish between 2D systems with different~$\tilde{\rho}$. To overcome the ensemble averaging, we developed an improved MRI approach to slice out a small subset of nearly identical 2D systems and measure their momentum distribution. In addition, we use matter wave focusing \cite{Amerongen2008} to more accurately identify the condensate, thereby reducing the measurement uncertainty in $f$.

We prepare a $N=2 \times 10^5$ atom $^{87}$Rb BEC~\cite{Lin2009}, with no discernable thermal component, in the $|\textrm{F}=1,m_\textrm{F}=1\rangle$ state in a 3D harmonic trap with measured trapping frequencies $\{ \nu_x,\nu_y,\nu_z\}=\{23.2(5),27.4(3),42.8(9)\}$ Hz. The trap arises from a combination of optical, magnetic and gravitational potentials. The BEC is confined at the intersection of a pair of 1064 nm laser beams, propagating along $\hat{x}$ and $\hat{y}$, with waists ($\exp(-2)$ radii) of about 55~$\mu$m. The BEC is~$620$~$\mu$m above the zero of a quadrupole magnetic field. At the center of the BEC the magnetic field is $B_0=193$ $\mu$T, corresponding to a Zeeman shift $g \mu_{\rm B} B_0/h=1.35$ MHz. The magnetic potential, nearly linear along $\hat{z}$ with a gradient of 2.180(4) kHz/$\mu$m, almost exactly cancels gravity and adds a harmonic anti-trapping potential in the $\hat{x}$-$\hat{y}$ plane for our $|\textrm{F}=1,m_\textrm{F}=1\rangle$ atoms.

We load the BEC into a 3D optical lattice at the intersection of three pairs of linearly polarized nearly counter propagating laser beams from a $\lambda=810$ nm Ti:Sapphire laser~\footnote{The vertical lattice beams intersect at $\theta$=164(1)$^\circ$, giving a lattice period of 421.3(2) nm. They differ in frequency from the beams in the $\hat{x}$-$\hat{y}$ plane by $\approx 160$ MHz, while the beams in the $\hat{x}$-$\hat{y}$ plane differ from each other by $2.824$ MHz and are cross polarized.}. These beams form independent~1D optical lattices along~$\hat{x}$,~$\hat{y}$ and~$\hat{z}$. The~$\hat{z}$ lattice is always set to a final depth of~24~$E_R$ and partitions the~3D~BEC into a set of~$\approx 60$~2D systems; the depth of the $\hat{x}$-$\hat{y}$ lattice ranges from~0~to~20~$E_R$ and determines the parameter~$U/t$. Together all confining potentials determine $\epsilon$. The recoil energy and momentum are~$E_R=\hbar^{2}k_R^{2}/2m=h\times3.4$~kHz and $k_R=2\pi/\lambda$. The lattices are turned on from zero intensity in~100~ms with a half-Gaussian intensity ramp (rms width of 37 ms). This time scale was chosen to be adiabatic with respect to interactions and all relevant single particle energy scales~\cite{Spielman2008,Gericke2006}. We measure lattice depth to within $\approx 2\%$ by pulsing on each lattice separately for~4 to~6~$\mu$s and observing the resulting atom diffraction~\cite{Ovchinnikov1998},\footnote{We calculate lattice depth from the pulse duration required for the $0^{th}$ and $1^{st}$ diffraction orders to be equally populated.}.

We implemented a MRI approach to selectively address a localized group of nearly identical 2D systems, as schematically illustrated in Fig.~\ref{Schem}a. A radio-frequency (rf) magnetic field $B_{\textrm{rf}}$, linearly polarized along $\hat x+\hat y$ transfers atoms from $m_\textrm{F}=1$ to $m_\textrm{F}=0$ and $m_\textrm{F}=-1$. We choose $B_{\textrm{rf}}$ to maximize the transfer into $m_\textrm{F}=0$ using a $400\us$ Blackman pulse (perfect transfer to $m_\textrm{F}=0$ is impossible for our 3 level system). The $2\kHz$ rms spectral width of this pulse, combined with the magnetic field gradient along~$\hat z$ gives a $0.9\micron$ rms spatial resolution ($\approx2$ lattice-sites), resulting in the extraction of 2D systems with up to 4000 atoms.

Following the rf pulse, the lattice potentials are ramped off with exponentially decreasing ramps ($400\us$ time constant) -- nearly adiabatic with respect to single particle energy scales in the optical lattice -- approximately mapping the occupied crystal momentum states in the lowest Brillouin zone to free momentum states~\cite{McKay2009}. At the same time, we remove the optical dipole trap in $<1\us$, and the magnetic field gradient along~$\hat{z}$ in~$\approx3\ms$; the atoms then expand for a $18.1\ms$ time-of-flight (TOF). During part of the TOF, a magnetic field gradient approximately along $\hat y$ quickly separates the three $m_\textrm{F}$ components. We then detect the final spatial distribution of all three components using resonant absorption imaging, which gives the approximate momentum distribution of each spin component separately. The $m_\textrm{F}=0$ distribution directly measures the momentum composition of the nearly identical 2D systems selected by the rf pulse, virtually eliminating the inhomogeneous averaging that is present in the $m_\textrm{F}=1$ distribution (see Fig.~\ref{Schem}c).

\begin{figure}[tb]
\begin{center}
\includegraphics[width=3.375in]{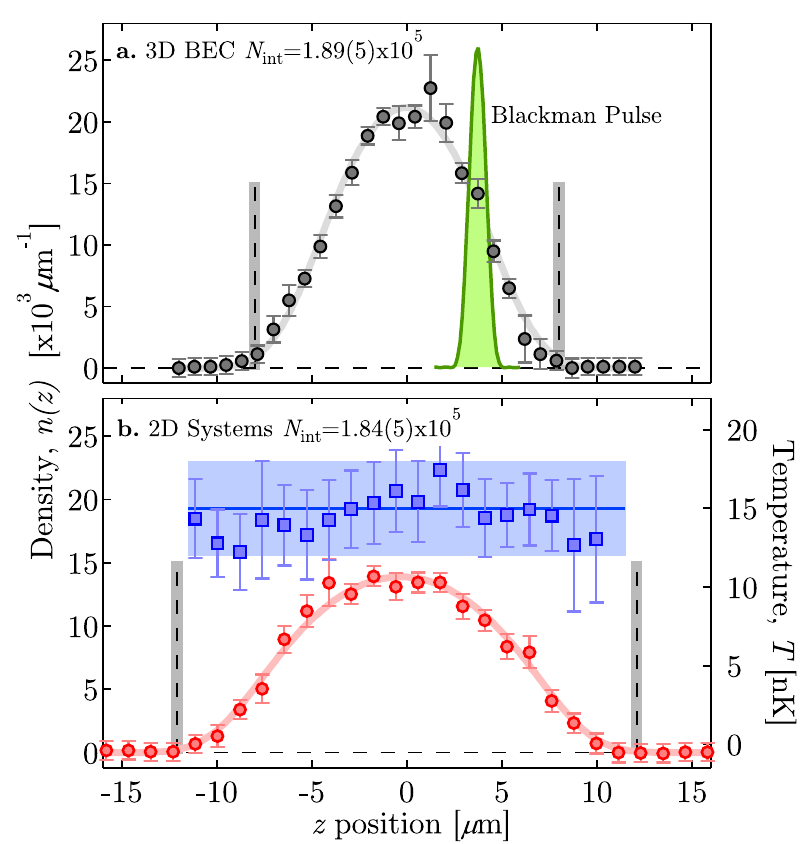}
\end{center}
\caption{Density profile $n(z)$ for: a) a 3D BEC and b) an ensemble of 2D systems. The atom number calculated from the $\textit{in situ}$ Thomas-Fermi radius $R_z=8.2(2)$ $\mu$m is $N_{\textrm{TF}}=1.8(4) \times 10^{5}$. The vertical dashed lines indicate the Thomas-Fermi radius from our fit. Continuous lines show a fit to the $\textit{in situ}$ 1D density profile $n(z)$. The temperature of the selected 2D systems (squares in b)) is displayed on the right axis, as a function of position along $\hat{z}$. On average $T=15(3)$~nK.}
\label{AtomNumber}
\end{figure}

Absorption images after TOF can differ from the \textit{in situ} momentum distributions for two primary reasons: a) interactions during TOF and b) finite TOF, here~$18.1\ms$. We mitigate each of these effects as follows:
a) The already weak interactions during TOF for the small number of atoms transferred into the $m_\textrm{F}=0$ state are further reduced by the rapid expansion along $\hat z$ after release from the tightly confining vertical lattice.
b) We used a matter-wave focusing technique -- a temporal atom lens -- that ``images'' the \textit{in situ} momentum distribution for a finite TOF ~\cite{Amerongen2008}. To focus the atoms we increase the harmonic trapping frequency by a factor of about 3, by linearly ramping the intensity of our 1064 nm dipole trap in $200\us$, and then holding for $400\us$ (during the rf pulse) just before TOF.

Our 3D BEC has a 56.9(4) $\mu$m Thomas-Fermi radius after TOF. When partitioned into an ensemble of 2D systems the radius decreases to $47.2(5)$ $\mu$m. For the extracted 2D systems the radius is $19.9(2)$ $\mu$m. Finally, figure~\ref{Schem}(c-d) illustrate the dramatic reduction to $10.5(2)\micron$ when both interactions and finite size effects are minimized.

We carefully calibrated atom number by measuring the {\it in situ} 1D density profile $n(z)$, of a 3D BEC using the MRI technique (see Fig.\ref{AtomNumber}a, circles). The Thomas-Fermi radius~$R_z=8.2(2) $~$\mu$m gives atom number~$N_{\textrm{TF}}=1.8(4) \times 10^{5}$; direct integration of~$n(z)$ gives~$N_{\textrm{int}}=1.89(5) \times 10^{5}$; measurement of absorption by all atoms after TOF gives~$N_{\textrm{abs}}=1.90(5) \times 10^{5}$. These measurements show that the combination of shot-to-shot number fluctuations and number measurement uncertainty is~$\sim$3\%. We confirm this by loading the BEC into the~1D optical lattice along~$\hat{z}$, and again measuring~$n(z)$. We find that the density profile expands along~$\hat{z}$ (Fig.~\ref{AtomNumber}b, circles) but the integrated atom number $N_{\textrm{int}}=1.84(5) \times 10^{5}$ remains constant. Figure \ref{AtomNumber}b also shows the measured temperature $T$ in a 1D optical lattice as a function of $z$ (squares). $T=15(3)$ nK is nearly uniform over all significantly occupied lattice sites, indicating that the 2D systems taken together are effectively in thermal equilibrium. Loaded into a shallow lattice this corresponds to a temperature~$k_\textrm{B}T=0.9(2) t$.

In our experiment we set $U/t$ using the $\hat{x}$-$\hat{y}$ lattice depth and~$\tilde{\rho}$ by rf-selecting 2D systems with the desired atom number from among the $\approx$ 60 available systems. As a result, each measured momentum distribution corresponds to a single point on the $U/t - \tilde{\rho}$ plane and we use the condensate fraction $f$ to distinguish between the SF and MI phases.

We experimentally define $f$ as the fraction of atoms in the sharp, focused feature in the momentum distribution. To remove the broad background, present due to thermal effects and quantum depletion, including atoms in the Mott phase, we fit to the thermal distribution of non-interacting classical particles in a 2D sinusoidal band~\cite{Spielman2008}. We smooth the fit function in a region within~$0.1$~$k_R$ of the edge of the Brillouin zone to account for non-adiabaticities in the lattice turn off near the band edge~\cite{McKay2009}. We exclude a disk with~0.16~$k_R$ radius around the condensate feature from the fit and identify the condensate as the atoms that remain within the disk after subtracting the fit.

\begin{figure}[tb]
\begin{center}
\includegraphics[width=3.375in]{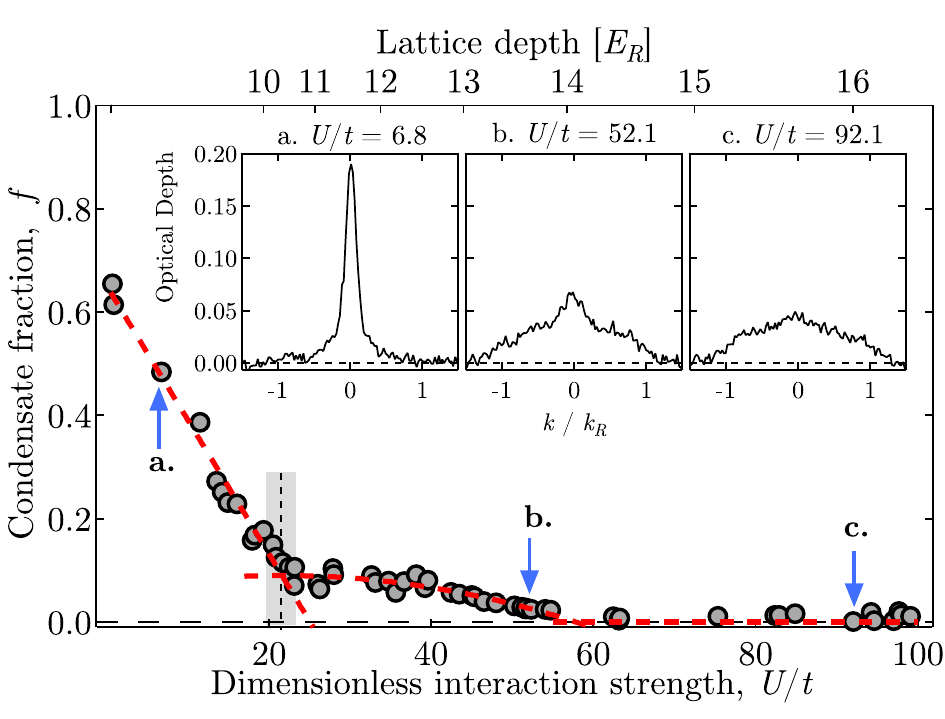}
\end{center}
\caption{Condensate fraction of a~2D Bose gas with~$N_{2D}\approx$~3500 atoms measured through the SF to MI transition; red dashed curves are the fits described in the text. Insets~(a-c) display the average momentum distribution $n(k)=[n_x(k)+n_y(k)]/2$ at different~$U/t$, where $n_x(k)$ is the momentum distribution integrated over $y$ and likewise for $n_y(k)$. We distinguish the formation of the first and second Mott regions at~$U/t=21(2)$ and~$U/t\approx$~55 respectively. For~$U/t>60$ the condensate fraction is indistinguishable from zero.}
\label{CondensateFractionFig}
\end{figure}

Figure \ref{CondensateFractionFig} shows the condensate fraction~$f$ as a function of~$U/t$ for 2D systems with $N_{2D}\approx3500$, $f$ up to~0.8, and an initial temperature~$T=0.9(2) t$, a factor of 2 lower than that reported in Ref.~\cite{Spielman2008} where $f \lesssim 0.4$ and $T \approx 2 t$. $f(U/t)$ exhibits three regions. For small $U/t$, $f$~decreases (fit to a line) until at $f \approx 0.12$ and $(U/t)_c=21(2)$ the slope changes markedly. We associate this feature with the first appearance of a MI and the subsequent decay (fit to a parabola) with the spatial growth of the MI regions. For $U/t>60$ the condensate fraction is indistinguishable from zero. The critical point for appearance of MI $(U/t)_c=21(2)$ is consistent with trapped system QMC~($(U/t)_c=20.5$ at $\tilde\rho=53$)~\cite{Rigol2009}. A similar analysis at $\tilde \rho \approx 20$ shows the first appearance of MI at $(U/t)_c=19(2)$, consistent with past measurements~($(U/t)_c=15.8(20)$)~\cite{Spielman2008}, homogeneous system QMC calculations~($(U/t)_c=16.5$)\cite{Krauth1991a,Elstner1999,Kato2007}, and trapped system QMC~($(U/t)_c=20$ at $\tilde \rho=20$)~\cite{Wessel2004,Rigol2009}.

Our measurements are summarized in Fig.~\ref{diagram}. We sampled about 1300 images with~$\tilde \rho$ up to~100 and~$U/t$ up to~100~\footnote{Our adiabaticity criterion for loading may break down when we enter into the regime of high $U/t$.} and for each image extracted the condensate fraction~$f$. The points are colored according to the side of the transition on which they are, light grey symbols denote the SF region and dark grey symbols show points with some MI. Red ovals mark the experimental phase transition boundary, their widths denote the uncertainties in the measurement of $\tilde \rho$ and $U/t$ where we identify the phase transition.

Figure \ref{diagram} corresponds with the predicted state diagram~\cite{Rigol2009}; the continuous curve shows the expected first appearance of a MI. The shift of this curve to larger~$U/t$ for increasing~$\tilde \rho$, reproduced by the data, goes beyond the LDA prediction where all phase transition lines are vertical, independent of~$\tilde \rho$, once MI shells form. The discrepancy for $\tilde \rho < 15$ is expected due to increased sensitivity to thermal effects at low density where the SF transition temperature is extremely low.

During the preparation of this paper we learned of a similar experimental technique applied to a~2D Bose system in the higher temperature BKT regime~\cite{Cornell2009}.
We appreciate enlightening conversations with C. Chin and J.K. Freericks; and we thank K. Mahmud and R. T. Scalettar for discussions and for sharing their QMC data (reproduced in Fig.~\ref{diagram}). This work was partially supported by ONR, DARPA's OLE program, and the NSF through the JQI Physics Frontier Center; K.J.G. thanks CONACYT and R.L.C. thanks the NIST/NRC program.

\bibliography{SF_MI_paper_bib}

\end{document}